\documentclass[%
 aip,
 amsmath,amssymb,
reprint,%
floatfix,
]{revtex4-1}

\usepackage{graphicx}% Include figure files
\usepackage[english]{babel}
\usepackage{dcolumn}% Align table columns on decimal point
\usepackage{bm}% bold math
\usepackage{xcolor}
\usepackage[utf8]{inputenc}
\usepackage[T1]{fontenc}
\usepackage{mathptmx}
\usepackage{eurosym}
\usepackage[normalem]{ulem}
\usepackage{multirow}
\usepackage{array}
\newcolumntype{P}[1]{>{\centering\arraybackslash}p{#1}}

\begin{document}

%\preprint{AIP/123-QED}

\title{Contagion-diffusion processes with recurrent mobility patterns of distinguishable agents}\bigskip
% Force line breaks with \\
\author{P. Valga\~n\'on}%
\affiliation{Departament of Condensed Matter Physics, University of Zaragoza, 50009 Zaragoza (Spain).}
\author{D. Soriano-Pa\~nos}%
\affiliation{Instituto Gulbenkian de Ci\^encia (IGC), 2780-156 Oeiras (Portugal)}
\affiliation{GOTHAM lab, Institute for Biocomputation and Physics of Complex Systems (BIFI), University of Zaragoza, 50018 Zaragoza (Spain).}
\author{A. Arenas}
\affiliation{Departament de Matem\'aticas i Enginyeria Inform\'atica, Universitat Rovira i Virgili, Tarragona (Spain).}
\author{J.G\'omez-Garde\~nes}
\email{gardenes@unizar.es}
\affiliation{Departament of Condensed Matter Physics, University of Zaragoza, 50009 Zaragoza (Spain).}
\affiliation{GOTHAM lab, Institute for Biocomputation and Physics of Complex Systems (BIFI), University of Zaragoza, 50018 Zaragoza (Spain).}
\affiliation{Center for Computational Social Science, University of Kobe, 657-8501 Kobe (Japan).} 
\date{\today}% It is always \today, today,

\begin{abstract}
 The analysis of contagion-diffusion processes in metapopulations is a powerful theoretical tool to study how mobility influences the spread of communicable diseases. Nevertheless, many metapopulation approaches use indistinguishable agents to alleviate analytical difficulties. 
 Here, we address the impact that recurrent mobility patterns, and the spatial distribution of distinguishable agents, have on the unfolding of epidemics in large urban areas. We incorporate the distinguishable nature of agents regarding both, their residence, and their usual destination. The proposed model allows both a fast computation of the spatio-temporal pattern of the epidemic trajectory and. the analytical calculation of the epidemic threshold. This threshold is found as the spectral radius of a mixing matrix encapsulating the residential distribution, and the specific commuting patterns of agents. We prove that the simplification of indistinguishable individuals overestimates the value of the epidemic threshold.
\end{abstract}

\maketitle

{\bf  Unraveling the influence that different aspects of human behavior have on how communicable diseases spread through populations is one of the most intriguing challenges in computational and theoretical epidemiology. Although it is now possible to include multiple types of human behavioral data for agent-based simulations, incorporating this entire arsenal of information into mathematical models to derive new analytical tools is a major challenge in epidemiology. In this article, we aim to go one step further on the road of increasing the realism of metapopulation-based epidemic models. In particular, we propose a theoretical framework that allows the inclusion of data on both the spatial distribution of populations and the distinction of agents according to their origin and destination. With this information at hand, we can derive the value of the epidemic threshold and its roots on the social mixing patterns that characterize the population under study providing, as a byproduct, a powerful tool to assess control strategies aimed at increasing its value under scenarios of epidemiological risk.
}

\section{Introduction}
\label{sec:introduction}

About 100 years ago, two physicians, Ronald Ross and Anderson G. McKendrick, and a chemist, William O. Kermack laid the foundations of epidemic modeling.  Mckendrik and Kermack formulated the celebrated SIR (Suseptible-Infectious-Recognized) compartmental model in 1927 \cite{SIR26,SIR27}, a framework that remains nowadays as the cornerstone of most theoretical works in epidemiology. Around the same time, in 1922, Lewis Fry Richardson \cite{Richardson} proposed a set of differential equations in an attempt to mathematically describe the evolution of the atmosphere and to go beyond qualitative weather forecasting to use quantitative and objective forms of prediction. 

Meteorological models quickly confirmed their practical usefulness as soon as they could be implemented in the first computing machines, seeing how their reliability improved as they were refined and both the quantity and quality of meteorological data increased. In contrast, despite theoretical advances in their refinement \cite{anderson1992infectious,keeling2011modeling}, the usefulness of epidemiological models remained limited for many decades, finding their fundamental utility in the qualitative understanding of the phenomena observed in different epidemic waves.  In this sense, the collection of epidemiological data for the validation and improvement of compartmental models was much more elusive than in the case of meteorology. In particular, apart from the biological features of the spreading pathogen, the main bottleneck for the development of reliable epidemic models was to accurately describe the human behavior underlying the observed infection patterns.

This barrier to the development of epidemiological models with predictive capacity was broken down with the advent of the 21st century, the internet era and the digitization of our daily lives.  The new digital era represents a paradigm shift for the study of human behavior on a large scale, allowing us, among other things, to access, and describe the skeleton of interactions through which infectious diseases are transmitted. Thus, in the last two decades, a great deal of effort has been invested in incorporating into epidemic models aspects such as the complexity of contact networks, patterns of human mobility at different geographic scales, and the time scales associated with human interactions \cite{review}.  Equipped with this information, theoretical, and computational tools can be used to analyze epidemiological problems, and develop forecasting frameworks that integrate both advanced epidemiological models and massive amounts of real demographic, mobility, and socioeconomic data at multiple scales\cite{eubank,gleam1,gleam2}. 

Most of these approaches rely on agent-based models that allow recreating synthetic populations mimicking the relevant social attributes that shape the unfolding of epidemic outbreaks. Although it is our most powerful tool in forecasting real epidemics and the evaluation of non-pharmaceutical interventions, agent-based simulations do not offer the possibility of obtaining transparent and analytical information about the importance of human behavior in the transmission of infectious diseases. To fill the gap between agent-based mechanistic simulations and theoretical frameworks, epidemic modeling relies on reaction-diffusion dynamics in metapopulations \cite{pion1,pion2}. This framework incorporates coarse-grained information of several features intervening in disease spreading such as the mobility patterns between different areas, the demographic distribution of a population and information about social mixing. 

In the last decade, the study of metapopulation dynamics has faced the challenge of approaching the realism of mechanistic simulations \cite{ball} by incorporating more and more aspects of human behavior and mobility \cite{barbosa2018human}. From the first works, including the complexity of human mobility networks \cite{colizza1,colizza2,colizza3,colizza4} the focus has been put on the recurrent nature of human mobility, being it of special importance in urban and regional scales \cite{commutes,prx,jtb,epjb}. Recently\cite{gomez2018critical}, we introduced a Markovian framework, the Mobility-Interaction-Return (MIR) model, that allowed the study of real populations incorporating the demographic distribution and the network of commutes. This approach revealed that these two aspects are essential to assess the advisability of contention measures based on the restriction of mobility. This approach has been further generalized to include networks with multiple types of mobility  \cite{Soriano-Panos2018,bosetti2020heterogeneity}, the study of vector-borne diseases \cite{PhysRevResearch.2.013312,anzo2019risk}, different permanence times on the destination \cite{JSTAT}, the heterogeneous of different contact patterns \cite{NJP}. Importantly, this Markovian framework has been used, after accounting for the particularities of SARS-CoV-2 transmission, to evaluate the evolution, and health systems impact, of COVID-19 in different countries \cite{COVID,Brazil,USA}. Likewise, the MIR model has been used to optimize resource allocation to control epidemics~\cite{zhu2021allocating} or to evaluate the role that individual awareness plays in hampering the spread of diseases~\cite{wang2021impacts}.
 
In this work, we go one step further in the formulation of the MIR epidemic model to better capture the recurrent mobility patterns of most human movements. To this aim, we get rid of one of the main hypotheses behind the former approach: indistinguishable agents, residents of a patch, according to their possible destinations. Including distinguishable agents allows us to analyze particular human commuting flows between different locations, and to identify those that are critical for the dissemination of infectious pathogens. This paves the path to inform about surgical interventions on the mobility patterns of a population to increase its resilience against the spread of a pathogen, in contrast to crude lockdowns spatially isolating one area. %We propose a MIR framework in which the set of residents of a subpopulation are divided into subsets of individuals having the same preferred destination. 
%After introducing and validating the formalism in Section~\ref{sec:Metapop}, we derive the expression for the epidemic threshold in Section~\ref{sec:threshold}. {\color{blue} In Section~\ref{sec:critical} we take advantage the distinguishable framework to unveil the most critical human flows for the unfolding of epidemics} and, in Section~\ref{sec:conc}, we round off by discussing the results and future research avenues.

\section{Results}

\subsection{Basic metapopulation framework}
\label{sec:Metapop}
Let us first introduce the basic MIR metapopulation framework that allows capturing the specific individual commuting patterns in generic populations in which agents display recurrent mobility patterns. In the following we will focus on the simple but paradigmatic Susceptible-Infected-Susceptible (SIS) compartmental model as the process underlying microscopic contagions. However, the formalism can be straightforward generalized to more sophisticated models, as is the case for the MIR metapopulation model with indistinguishable agents \cite{Soriano-Panos2018,PhysRevResearch.2.013312,JSTAT,COVID}.

The SIS model describes a process in which a person who is in a susceptible state, upon contact with an infected person, becomes infected with probability $\lambda$. At the same time, an infected person recovers with probability $\mu$ and, at variance with the SIR model, becomes susceptible again. When writing the equations for a single population of $n$ interacting agents, one typically considers the fraction of infected individuals in the population $\rho(t)$. The evolution of this variable can be written considering a mean-field approximation, {\em i.e.} assuming that all individuals are equivalent and have a homogeneous probability of interacting with each other. This way, assuming that each agent makes $\langle k\rangle$ contacts at time $t$ and the existence of a fraction $\rho(t)$ of infected individuals at time $t$, the probability that a susceptible agent gets infected at this time reads: 
\begin{equation}
P(t) = 1 - (1 - \lambda \cdot \rho(t))^{\langle k\rangle}\;.
\label{SIS1}
\end{equation}
The expression of this probability allows us to write the mean-field evolution for a single population of $n$ agents. Considering a time-discrete version, the fraction of infected individuals at time $t+1$ is given by:
\begin{equation}
\rho (t+1) = (1 - \mu) \rho(t) + P(t) \cdot (1 - \rho (t))\;.
\end{equation}

Although simple, the SIS model captures the most relevant feature of an epidemic process: the epidemic threshold $\lambda_c$. The epidemic threshold is the minimum value of $\lambda$ yielding an epidemic scenario in which the epidemic is not extinguished but keeps circulating from one individual to another. For the case of the mean-field SIS model the epidemic threshold can be easily derived, $\lambda_c = \frac{\mu}{\langle k \rangle}$, and implies that when $\lambda>\lambda_c$ the system reaches a steady-state in which the stationary value of $\rho(t)$ is constant and non-zero since new-infections are balanced by the recovery of infected agents. 

In epidemiological terms, the epidemic threshold $\lambda_c$ is closely related to the so-called basic reproduction number $R_0$, defined as the number of secondary infections a single infectious individual would make in a population of fully susceptible agents. The basic reproduction number in the mean-field SIS model is $R_0=\lambda \langle k \rangle/\mu$. Thus, when the infectivity per contact $\lambda=\lambda_c=\frac{\mu}{\langle k \rangle}$ the system has a reproduction number $R_0=1$, meaning that, on average, an infected agent makes $1$ new infection during its infectious period $\mu^{-1}$. Obviously when $\lambda>\lambda_c$ the corresponding reproduction number is $R_0>1$ and corresponds to an epidemic regime.

The mean-field SIS model contains many simplifications and can be improved in many ways to become a more realistic framework. One of its main assumptions is that the population under study is completely isolated. However, as many recent real epidemics reveal, the main aspect behind the explosion of localized outbreaks into epidemic (or even pandemic) scenarios is the high mobility of individuals between different populations. To incorporate this important feature of real epidemics into any compartmental model as the SIS one draws on the metapopulation formalism.  

A basic metapopulation model describes the contacts between individuals at two different scales by dividing the total population into subpopulations of different sizes. In each subpopulation, individuals interact and have contact with each other, facilitating the contagion of the pathogen. In turn, agents are allowed to move and visit different subpopulations, thus favoring the transmission of the pathogen to disease-free regions.  
When defining the metapopulation framework one has to set the microscopic contagion dynamics happening within the patches (here the SIS model) and the type of diffusion that better describes the mobility of agents. To capture the typical mobility patterns at urban or regional scales, one has to take into account the recurrent nature of human mobility at these scales  \cite{barbosa2018human}. As motivated above, in \cite{gomez2018critical} we introduced the MIR model in which each individual has an associated subpopulation (the residence) and comprises a three-stage process in every time step. The first stage comprises the diffusion step in which each individual decides, with probability $p_d$, to travel to a destination or to stay (with probability $1-p_d$) in its residential patch. Second, the reaction process in which contacts occur between individuals being in the same subpopulation at time $t$. Finally, the return to the original subpopulations of those agents that decided move in the first stage of the MIR sequence. In this final stage, we also take into account, as introduced by Granell and Mucha\cite{Granell}, that agents make the second round of interactions at the household level. This way interactions can be split into those made during the day ($D$) and those made at night ($N$) with their corresponding housemates. 

The recurrent mobility of agents is typically provided by Origin-Destination (OD) matrices. Thus for a population divided into $N$ subpopulations one has an $N\times N$ OD matrix whose $(i,j)$-entry contains the number $n_{ij}$ of agents with residence in $i$ that typically perform daily commutes to patch $j$. This matrix can be viewed as a directed (as commuting patterns between two patches are not symmetric) and weighted network connecting the collection of $N$ patches. 

Once equipped with the OD matrix characterizing the mobility flows of a metapopulation and the knowledge about the census of each subpopulation ($\{n_i\}$), one can tackle the formulation of the dynamical equations that rule the evolution of a metapopulation. To this aim, in \cite{gomez2018critical} the authors assign to each node $i$ a probability that an individual is infected $\rho_i(t)$, which means that we can define the state of the metapopulation via the vector $\vec{\bf \rho}^T=(\rho_1(t), ..., \rho_N(t))$ that can be used to compute the local prevalences at each patch $i$ as $n_i \cdot \rho_i(t)$. The time evolution of the variables $\rho_i(t)$ defining the epidemic state of each patch can be written as the following time-discrete Markovian chain:
\begin{equation}
\rho_i(t+1) = (1-\mu)\rho_i(t) + (1-\rho_i(t))\Pi_i(t)\;,\label{MIR1}
\end{equation}
where $\Pi_i(t)$ is the probability that an individual with residence in patch $i$ becomes infected at time $t$:
\begin{multline}
\Pi_i(t) = (1 - p_d) \left(P_i^D(t) + (1 - P_i^D(t))P_i^N(t)\right) + \\
    p_d \left(\sum_{j=1}^N R_{ij} \left( P_j^D(t) + (1 - P_j^D(t))P_i^N(t)\right) \right)\;.
\label{MIR2}
\end{multline}
The former expression contains in its {\em r.h.s.} two terms accounting for either the infection at its residence $i$ and the infection at a patch $j$ that, in general, is different from $i$. These two terms are weighted by $(1-p_d)$ and $p_d$ respectively since, as introduced above,  $p_d$ is the probability that agents move in the metapopulation, {\em i.e.} a control parameter that allows us to tune the level of spatial confinement in the population. Besides, the second infection probability in Eq.~(\ref{MIR2}) makes use of matrix ${\bf R}$ in which each entry $R_{ij}$ is defined as the probability that a moving agent with residence in $i$ chooses patch $j$ as the destination. The matrix ${\bf  R}$ is constructed directly from the OD matrix, $R_{ij}=n_{ij}/\sum_{l}n_{il}$, and satisfies the normalization condition of a row stochastic matrix: $\sum_{j=1}^{N} R_{ij} = 1$ $\forall \ i$.

Finally, both terms in the r.h.s of Eq.~(\ref{MIR2}) contain two sets of probabilities, $\{P_i^D(t)\}$ and $\{P_i^N(t)\}$, that account for the probabilities of getting infected being placed at a patch $i$ during the day (D) and of contracting the disease at the household (N) placed in patch $i$ respectively. In both cases, these probabilities adopt a similar form to Eq.~(\ref{SIS1}). Namely:
\begin{eqnarray}
P_i^D(t) &=& 1 - \left( 1 - \lambda\dfrac{I_i^{eff}(t)}{n_i^{eff}} \right)^{z^D f_i} \;,\label{MIR4_1}\\
P_i^N(t) &=& 1 - \left( 1 - \lambda \rho_i(t) \right)^{z^N \sigma_i} \;.
\label{MIR4_2}
\end{eqnarray}
Eq.~(\ref{MIR4_1}) incorporates the overall population $n_i^{eff}$ and the effective number of infected individuals $I_i^{eff}$ located at patch $i$ after the Movement stage. Under the assumptions of the model, both quantities read:
\begin{eqnarray}
n_i^{eff} &=& \sum_{j=1}^N \left( (1-p_d)\delta_{ij} + p_d R_{ji} \right) n_j\label{MIR5}\;,\\
I_i^{eff} &=& \sum_{j=1}^N \left( (1-p_d)\delta_{ij} + p_d R_{ji} \right) n_j \rho_j(t) \;.
\end{eqnarray}
In contrast, Eq.~(\ref{MIR4_2}) only considers the fraction of infected individuals residing in patch $i$, as it governs the probability of contagion with the members from the household. Note that, in both expressions, we have denoted the number of contacts made in patch $i$ during the D and N cycles by $z^D f_i$ and $z^N \sigma_i$ respectively. On one hand, the number of contacts during D is proportional to the patch population density:
\begin{equation}
f_i = \dfrac{n_i^{eff}}{a_i}\;,
\label{MIR5(3)}
\end{equation}
while $z_D$ is a scaling factor that ensures that the average number of contacts during D across the entire metapopulation remains equal to $\langle k^D\rangle$:
\begin{equation}
z^D = \dfrac{\sum_{i=1}^N n_i^{eff}\langle k^D \rangle}{\sum_{i=1}^N n_i^{eff}f_i} \;.
\label{MIR5_1}\\
\end{equation}
On the other hand, for the N cycle, the average number of contacts is proportional to the average housemates at given patch $i$, $\sigma_i$ and $z_N$ ensures that the average number of contacts across the entire metapopulation is $\langle k^N \rangle$:
\begin{equation}
z^N = \dfrac{\sum_{i=1}^N n_i\langle k^N \rangle}{\sum_{i=1}^N n_i\sigma_i}\;.
\label{MIR5_2}
\end{equation}

% In these two probabilities we include the effective number of individuals that are placed in a patch $i$ after the Movement stage has taken place:
% \begin{equation}
% n_i^{eff} = \sum_{j=1}^N \left( (1-p_d)\delta_{ij} + p_d R_{ji} \right) n_j\;,
% \label{MIR5}
% \end{equation}
% the effective number of infectious individuals placed in a patch $i$ after the Movement stage:
% \begin{equation}
% I_i^{eff} = \sum_{j=1}^N \left( (1-p_d)\delta_{ij} + p_d R_{ji} \right) n_j \rho_j(t) \;,
% \label{MIR5(2)}
% \end{equation}
%and the average number of contacts at a given patch $i$ during the D cycle, $z^D f_i$, and at a household placed in patch $i$, $z^N \sigma_i$. 

\begin{figure*}[t!]
	\centering
	\includegraphics[width=0.85\linewidth]{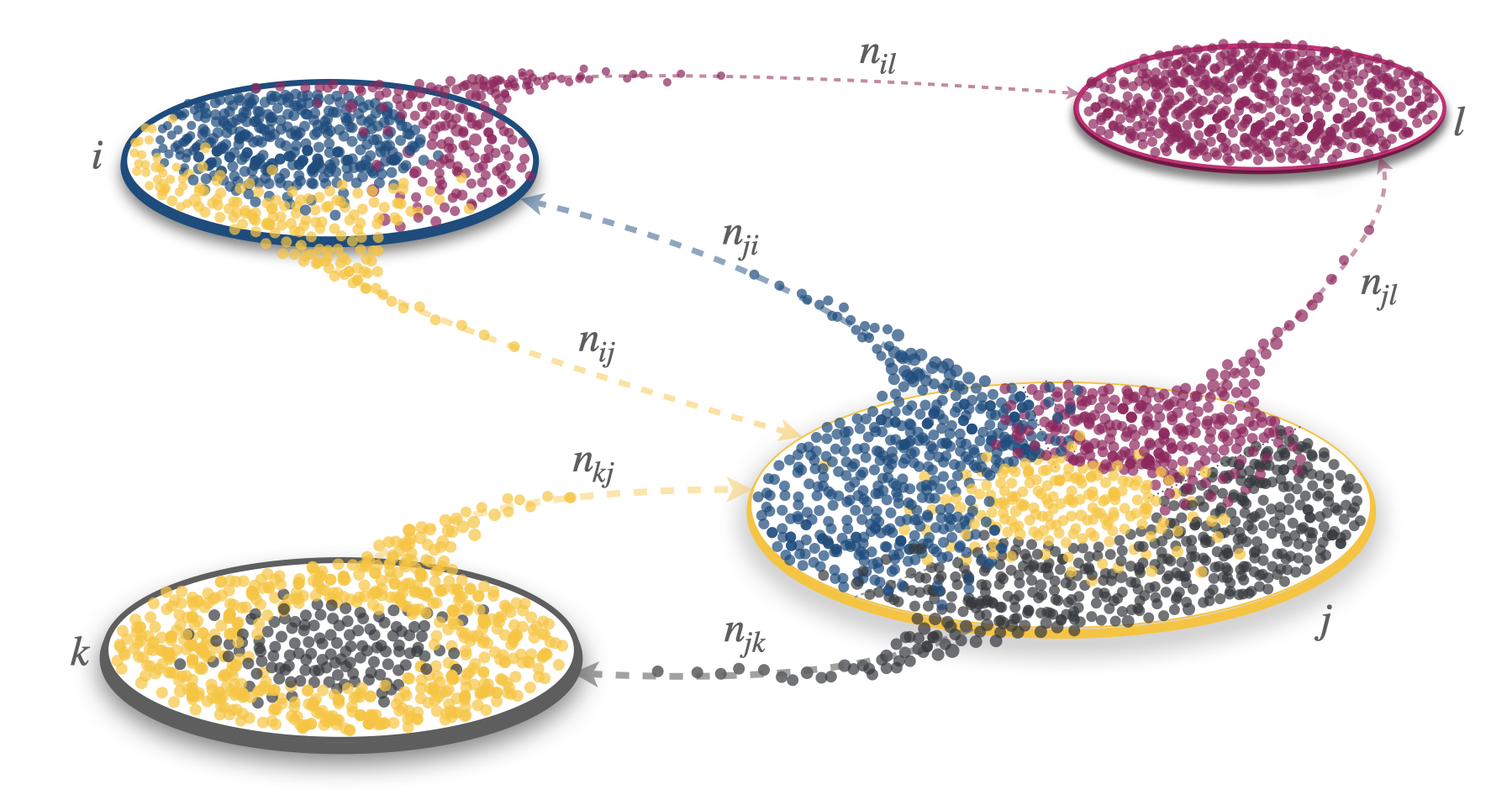}
	\caption{Schematic representation of a toy metapopulation network with distinguishable agents. The population is divided into $4$ interconnected subpopulations denoted as $i$, $j$, $k$ and $l$. In each subpopulation the agents are divided (colored) according to their destinations. This way the number of agents that travel from $i$ to $j$ correspond to a subset of size $n_{ij}$. Note that each patch also contains a subset of agents whose destination is the same as the residence and thus are colored according to the color of the corresponding patch.}
	\label{fig:scheme}
\end{figure*}

As shown in \cite{gomez2018critical}, the former formulation nicely agrees with the results obtained in mechanistic simulations of general metapopulations, such as cities or regions, for which the local census and the daily commuting trips between neighborhoods or municipalities are known. In addition, the mathematical formulation allows deriving an analytical expression for the epidemic threshold:
\begin{equation} 
\label{eq:umbralIndist}
\lambda_c = \dfrac{\mu}{\Lambda_{max}(\bm{M})}\;,
\end{equation}
where $\Lambda_{max}(\bm{M})$ is the maximum eigenvalue of the {\em mixing matrix} ${\bf M}$ defined as:
\begin{multline}
	M_{ij} = \left[ \left( (1 - p_d)^2 \frac{z^D f_i}{n_i^{eff}} + \frac{z^N \sigma_i}{n_i} \right) \delta_{ij} +\right.\\
	\left. p_d(1 - p_d)\left( R_{ji}\frac{z^D f_i}{n_i^{eff}} + R_{ij} \frac{z^D f_j}{n_j^{eff}}\right) + \right.\\
	\left. p_d^2 \sum_{l=1}^N R_{il}R_{jl} \frac{z^D f_l}{n_l^{eff}}\right] n_j\;.\;\;\;\;\;\;\;\;\;\;\;\;\;\;\;\;\;\;\;\;\;\;\;\;\;\;\;\;\;\;\;\;\;\;\;\;\;
	\label{mixing_indist}
\end{multline}
Each term, $M_{ij}$, of ${\bf M}$ contains the three elementary processes by which agents from patches $i$ and $j$ can interact. The main novelty of this formalism shows up when analyzing the dependence of the epidemic threshold with the degree of mobility $p_d$ since, counter intuitively, it is shown that mobility can help to decrease the epidemic prevalence and hence increase the epidemic threshold. This result is the product of incorporating both the heterogeneous demographic distribution of real cities or regions and the commuting nature of human mobility.

\subsection{Incorporating the distinguishable nature of individuals} \label{sec:distinguibles}

The original MIR model and its subsequent refinements assume that all individuals belonging to the same subpopulation $i$ are equivalent, {\em i.e.} they explore all the possible destinations that are connected to this patch $i$ according to the data contained in the OD matrix and captured in the right stochastic matrix ${\bf R}$. However, this assumption neglects that each agent has its patterns of movement.

To overcome this limitation, here we consider that the agents with residence in a patch $i$ are distinguishable according to their preferred destination. As a result, we cannot assume that there will be the same proportion $\rho_i(t)$ of infected people living in $i$ among the subsets of individuals traveling to different destinations. Thus, we have to further divide the set of individuals having residence in $i$ and consider a new set of variables $\{\rho_{ij}(t)\}$ that account for the probability that an individual with residence in patch $i$ whose usual destination is node $j$ is infected at time $t$. 
%Note that in the indistinguishable formulation the epidemic prevalence of the metapopulation was characterized by a vector of $N$ components, $\vec{\rho}$ whereas in the distinguishable framework we have to keep track of a number of variables equal to the number $L$ of commuting links in the metapopulation. 
This way, the number of variables raises from $N$ (indistinguishable case) to $L\leq N^2$ (indistinguishable case), being $L$ the total number of non-zero entries in the OD matrix. In the most general scenario, the connectivity through commuting patterns between the $N$ patches is characterized by an $N^2\times N^2$ matrix ${\bf N}$, whose $(i,j)$-entry, $n_{ij}$, accounts for the number of commuters with residence in $i$ whose usual destination is patch $j$. Let us note that for each patch $i$ there is a subset of individuals whose usual commuting destination is node $i$, {\em i.e.} in general $n_{ii}\neq 0$. Naturally, these variables must satisfy that $\sum\limits_{j=1}^N n_{ij} = n_i\ \forall i$. This distinguishable framework is illustrated in Fig.~\ref{fig:scheme}, where we show a schematic simple metapopulation of $4$ patches in which the agents are distinguished (colored) according to their usual commuting destination. 
%Also, here we will make use of the information provided by the OD matrix, {\em i.e.} without the need of the stochastic matrix ${\bf R}$ and working directly with the total number of individuals that typically travel from $i$ to $j$, $n_{ij}$.

To formulate the Markovian equations we consider again that at each time step susceptible agents can be infected during the two stages of the MIR sequence. First, during the day at stage I, when agents interact either at their destination with probability $p_d$ or at their residential patch with probability $(1-p_d)$, and second, at the household level at stage R. The infection probability in the first case, when an agent is placed at patch $i$:
\begin{equation}
P_i^D = 1 - \left( 1 - \lambda\dfrac{I_i^{eff}(t) }{n_i^{eff}}\right)^{z^D f_i}\;,
\label{MIRD1}
\end{equation}
where $n_i^{eff}$ is, as in Eq.~(\ref{MIR5}), the effective number of individuals that are placed in a patch $i$ that, for the distinguishable case, reads:
\begin{equation}
n_i^{eff} = p_d \sum_{j=1}^N n_{ji} + (1 - p_d)\sum_{j=1}^Nn_{ij}\;.
\label{MIRD2}
\end{equation}
The r.h.s. of the former expression contains the number of people that decide to travel to $i$ at the $M$ stage from any patch $j$ while the second contains accounts for all the commuters departing from patch $i$ that decide not to travel. Following the same rationale is easy to write the effective number of infected individuals that visit a patch $i$ once the movement stage has taken place:
\begin{equation}
I_i^{eff}(t) = p_d \sum_{j=1}^N n_{ji} \rho_{ji}(t) + (1 - p_d)\sum_{j=1}^N n_{ij} \rho_{ij}(t)\;.
\label{MIRD3}
\end{equation}
With the former two expressions, $I_i^{eff}$ and $n_i^{eff}$, the probability that an agent at patch $i$ is infectious, $I_{i}^{eff}/n_{i}^{eff}$ can be constructed and used in Eq.~(\ref{MIRD1}). In close analogy with Eq.~(\ref{MIR4_1}) for the indistinguishable case we consider that the number of contacts at patch $i$ is proportional to the effective population density:
\begin{equation}
f_i = \dfrac{n_i^{eff}}{a_i}\;,
\label{MIRD4}
\end{equation}
scaled by:
\begin{equation}
z^D = \dfrac{\sum_{i=1}^N n_i^{eff}\langle k^D \rangle}{\sum_{i=1}^N n_i^{eff}f_i} \;, 
\label{MIRD5}
\end{equation}
to ensure that the average number of contacts in the entire metapopulation at the I stage (D period) is equal to $\langle k^D \rangle$.

For the probability of infection at the R stage (N period) we have:
\begin{equation}
P_i^N = 1 - \left( 1 - \lambda \frac{\displaystyle\sum_{j=1}^N n_{ij}\rho_{ij}(t)}{\displaystyle\sum_{j=1}^N n_{ij}} \right)^{z^N \sigma_i}\;,
\label{MIRD6}
\end{equation}
where the fraction in the r.h.s has as numerator the expected number of infected individuals with residence at patch $i$ while the denominator is the population of patch $i$. Thus, this fraction encodes the probability that a resident at patch $i$ is infectious. As in Eq.~(\ref{MIR4_2}) we consider that in the R stage (period N) interactions are restricted to the household level so that they are given by $z_N\sigma_i$, where $\sigma_i$ is the average housemate number of households at patch $i$ and $z^N$ reads: 
\begin{equation}
z^N = \dfrac{\sum_{i=1}^N n_i\langle k^N \rangle}{\sum_{i=1}^N n_i\sigma_i}\;,
\label{MIRD7}
\end{equation}
so that the average household contacts across the whole metapopulation during period N is $\langle k^N \rangle$.

Equipped with the infection probabilities during periods D and N, i.e. Eqs.~(\ref{MIRD1}) and (\ref{MIRD6}), we can construct the infection probability for a resident of patch $i$ having patch $j$ as her usual commuting destination:
\begin{equation}
\label{MIRD8}
\Pi_{ij} = (1 - p_d)\left( P_i^D + (1 - P_i^D)P_i^N\right) + p_d\left( P_j^D + (1 - P_j^D)P_i^N\right)\;.
\end{equation}
The former probability allows us to write the dynamical evolution for the fraction of infected agents with residence in $i$ and destination $j$:
\begin{equation}\label{MIRD9}
\rho_{ij}(t+1) = (1-\mu)\rho_{ij}(t) + (1 - \rho_{ij}(t))\Pi_{ij}\;,
\end{equation}
where the first term accounts for the fraction of infectious commuters between $i$ and $j$ at time $t$ that do not recover whereas the second term adds the new infections of susceptible $(i,j)$-commuters that take place at time $t$ both at the residence $i$ and the destination $j$. The former expression takes part of a set of $L$ equations that can be solved by iterating a given initial condition $\{\rho_{ij}(0)\}$. 

\begin{figure}[t!]
    \centering
		\label{subfig:Verificacionnewyork}
		\includegraphics[width=1.1\linewidth]{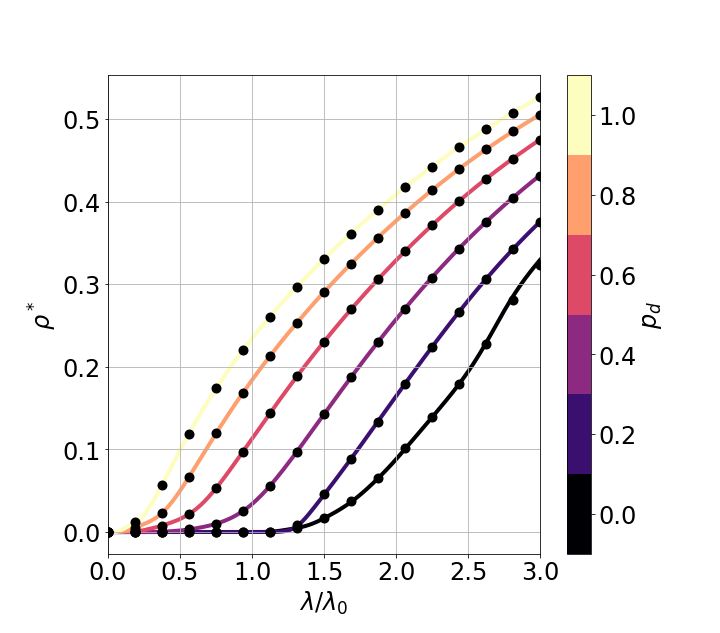}
		\label{subfig:Verificacionboston}
		\includegraphics[width=1.1\linewidth]{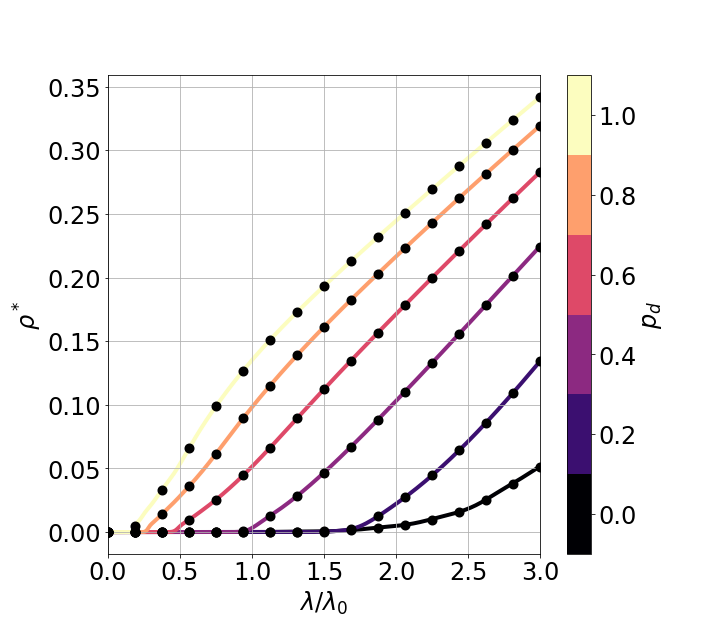}
	\caption{Validation of the Markovian equations for the cities of $(a)$ New York and $(b)$ Boston. On the vertical axis we represent the fraction of infected $\rho^*$ in the stationary state for the whole population, and on the horizontal axis the parameter $\lambda$ normalized to the epidemic threshold for $p_d = 0$: $\lambda_0=\lambda_c(p_d=0)$. Each panel shows different epidemic curves $\rho^*(\lambda/\lambda_0)$ for different values of $p_d$ obtained by iterating the Markovian equations. In its turn, points represent the  average of $\rho^*$ obtained after $100$ realizations of the mechanistic (Monte Carlo) simulations for each value of $\lambda$ and $p_d$ (note that error bars are smaller than points). In all the cases the recovery probability is set to $\mu=0.2$, the average contacts during the D cycle are $\langle k^D\rangle=8$, and those that take place at the N one are set to $\langle k^D\rangle=3$.}
	\label{fig:graficaMC}
\end{figure}

\subsection{Model Validation}

To validate the set of Markovian equations given by Eq.~(\ref{MIRD9}) we now construct real metapopulations incorporating the demographic distribution and the commuting flows of the population of two different core-based statistical areas (cbsa) in the United States (see Table~\ref{tab:Table1} for details); namely New York-Newark-Jersey City and Boston-Cambridge-Newton. For brevity, we will refer to them in what follows as New York and Boston respectively. For each cbsa, the patches represent the different zip codes, whereas the corresponding OD matrices have been obtained from surveys capturing the population moving daily from one area to another. Data about the commuting flows \cite{UScomm}, the distribution of the population across patches \cite{UScensus} and the area of each patch \cite{USarea} are publicly available. A summary of the main attributes for every metapopulation studied throughout the paper is available in Table \ref{tab:Table1}. \\

\begin{table}[t!]
\centering
\begin{tabular}{|l|c|c|c|c|}
\hline
\textbf{Name} & \multicolumn{1}{l|}{\textbf{Population}} & \multicolumn{1}{l|}{\textbf{Patches}} & \multicolumn{1}{l|}{\textbf{Links}} & \multicolumn{1}{l|}{\textbf{Density $(sq. miles^{-1})$}} \\ \hline
New York      & 12 423 494                               & 498                                   & 148 001                             & 5 084                                                         \\ \hline
Boston        & 4 146 213                                & 232                                   & 42 064                              & 1 704                                                         \\ \hline
Austin        & 1 775 659                                & 98                                    & 7 281                               & 306                                                           \\ \hline
Miami         & 5 590 269                                & 186                                   & 31 790                              & 1 912                                                         \\ \hline
Detroit       & 4 475 286                                & 232                                   & 40 220                              & 964                                                           \\ \hline
Seattle       & 3 502 087                                & 176                                   & 22 375                              & 565                                                           \\ \hline
\end{tabular}
\caption{Main characteristics of the $6$ US cbsa studied as metapopulations. Namely, the total population, the number of patches and links that compose each metapopulation, and the average population density of the patches.}
\label{tab:Table1}
\end{table}

Once the census and mobility data have been translated into a metapopulation we first perform mechanistic simulations by considering the $n$ agents of each cbsa and  simulating the microscopic dynamics corresponding to both individual displacements and also the pairwise interactions of susceptible and infectious agents that give rise to contagions of the former. We carry these simulations by considering different values of the infectivity $\lambda$ and the mobility parameter $p_d$ while keeping the recovery probability $\mu = 0.2$. For each simulation, we let the system evolve for a reasonable time of $T$ days (here the natural time unit imposed by the commuting data) and compute the stationary value for the fraction of infected individuals, $\rho^*$. Since  mechanistic simulations are stochastic, for each value of $\lambda$ and $p_d$, we consider the average value of $\rho^*$ obtained from $100$ realizations of the mechanistic simulation with different initial conditions for the infectious seeds. 

The results obtained through mechanistic simulations are confronted with those obtained by iterating the Markovian equations by comparing the stationary value for the fraction infectious individuals. In the case of the Markovian equations, the steady fraction of infectious agents, {\em i.e.} the average prevalence of the diseases, is calculated as:
\begin{equation}
\rho^*=\frac{1}{n}\sum_{i,j=1}^N n_{ij}\rho_{ij}^*
\end{equation}
where the values $\rho_{ij}^*$ are the stationary values of the entries of vector ${\bf {\rho}}$ and $n$ is the total population in the system. Note that, since Markovian dynamics is deterministic and there are neither sources or sinks in the compartmental model, no realizations are needed, thus allowing a fast calculation of epidemic curves $\rho^*(\lambda)$.  

The results obtained by both methods are shown in Figure \ref{fig:graficaMC}. In both panels, the fraction of infected individuals is represented as a function of the contagion probability $\rho(\lambda)$ for different values of $p_d$. The infectivity value is re-scaled by a value $\lambda_0$, which is the epidemic threshold $\lambda_c$ when $p_d=0$, {\em i.e.} $\lambda_0=\lambda_c(p_d=0)$. The value $\lambda_0$ is that of the SIS model in the most vulnerable patch, {\em i.e.} $\lambda_0 = \mu/\max({z^N\sigma_i + z^D f_i})$. From the plots it is clear that the agreement between the Markovian solution (curves) and the results from mechanistic simulations (points) is excellent, thus confirming the validity of the Markovian equations~(\ref{MIRD9}).

Remarkably, the epidemic curves for the two cities show the so-called epidemic detriment driven by mobility, {\em i.e.} the increase of the epidemic threshold $\lambda_c$ for values $p_d>0$ compared to the case $p_d=0$. Thus, the distinguishable MIR framework preserves the main result derived by the indistinguishable one \cite{gomez2018critical}, as it is rooted on the heterogeneous distribution of the population and the flows across patches. This counter-intuitive phenomenon can be easily explained considering the  case $p_d = 0$ and $\lambda\gtrsim \lambda_0$. In this case, infectious agents are concentrated in the most vulnerable areas while the rest of the patches are disease-free subpopulations. By increasing $p_d$ we spread the carriers of the disease to other areas with lower infection risk where, for the same value of $\lambda\gtrsim \lambda_0$, will not spread the pathogen. Besides, those individuals moving from the least to the most exposed areas are generally healthy. Thus, for $p_d\gtrsim 0$ and $\lambda\gtrsim \lambda_0$, the infected individuals will recover without making secondary infections in small size patches, while the localized outbreak in the largest population patch dies out.

\subsection{Epidemic threshold}
\label{sec:threshold}

Once the model has been formulated and validated, obtaining as a byproduct the confirmation that the epidemic detriment driven by mobility remains in the distinguishable MIR formulation, we now tackle the analysis of the epidemic threshold $\lambda_c$. To this aim we focus on finding the stationary state of the SIS dynamics and thus assume that all the variables are time-independent: $\rho_{ij}(t+1)=\rho_{ij}(t)\equiv\rho_{ij}^*$ $\forall i,j$. In this stationary regime Eq.~(\ref{MIRD9}) transforms into: 
\begin{equation}
\label{eq:Markov3}
\mu\rho_{ij}^* =(1 - \rho_{ij}^*)\Pi_{ij}(\vec{\rho^*})\;.
\end{equation}
Solving this equation by numerical means allows us to find the stationary value $\rho^*$ used in Fig.~(\ref{fig:graficaMC}). However, since our focus is those solutions close enough to $\lambda_c$, we are interested in a very particular situation that consists of arbitrarily small local prevalences: $\rho_{ij}^*\equiv \varepsilon_{ij} \ll 1$ $\forall i,j$. This way we can linearize Eq.~(\ref{eq:Markov3}). 

Since the function $\Pi_{ij}(\vec{\rho^*})$ in Eq. (\ref{eq:Markov3}) is highly nonlinear we start by considering the two sets of contagion probabilities $\{P^{D}_i(\vec{\rho}^*)\}$ and $\{P^{N}_i(\vec{\rho^*})\}$ and linearize each of them as: 
\begin{eqnarray}
P_i^D &\simeq& p_d \lambda \dfrac{z^D f_i}{n_i^{eff}}\sum_{j=1}^N n_{ji}\varepsilon_{ji} + \lambda (1 - p_d) \dfrac{z^D f_i}{n_i^{eff}} \sum_{j=1}^N n_{ij} \varepsilon_{ij}\;, \\
P_i^N &\simeq& p_d \lambda \dfrac{z^N \sigma_i}{n_i} \sum_{j=1}^N n_{ij}\varepsilon_{ij}\;.
\end{eqnarray}
Next we insert the former expressions into Eq.~(\ref{MIRD8}) to obtain the linearized version of $\Pi_{ij}(\vec{\rho^*})$ as:
\begin{multline}
\Pi_{ij}(\vec{\epsilon}) = \lambda \sum_{k,l=1}^N \left[ (1 - p_d) p_d \frac{z^D f_k}{n_k^{eff}}n_{lk}\delta_{ik}\right.\\
\left.+ (1 - p_d)^2 \frac{z^D f_l}{n_l^{eff}} n_{lk}\delta_{il} \right.\\
 \left.+p_d^2 \frac{z^D f_k}{n_k^{eff}}n_{lk}\delta_{jk}\right.\\
 \left.+p_d(1 - p_d)\frac{z^D f_l}{n_l^{eff}}n_{lk}\delta_{jl}\right.\\
 \left. + \frac{z^N \sigma_l}{n_l}n_{lk}\delta_{il} \right] \lambda \varepsilon_{lk}\;\;\;\;\;\;\;\;\;\;\;\;\;\;\;\;\;\;\;\;\;\;
\label{pi_ij}
\end{multline}
Finally, inserting this formula into Eq.~(\ref{eq:Markov3}) and neglecting nonlinear terms in $\epsilon_i$, we obtain the following set of linear equations for the stationary prevalence $\vec{\epsilon}$:
\begin{equation}
\label{eq:Linealizada}
\mu \epsilon_{ij} \simeq \Pi_{ij}(\vec{\epsilon}^*) = \lambda \sum_{k,l=1}^N M_{jk}^{il} \varepsilon_{lk}
\end{equation}
where, for convenience, we have written the expression in Eq.~(\ref{pi_ij}) as the product of the prevalence vector $\vec{\epsilon}$ by a matrix {\bf M} whose terms are defined as:
\begin{multline}
\label{eq:mixing}
M^{il}_{jk}=(1 - p_d) p_d \frac{z^D f_k}{n_k^{eff}}n_{lk}\delta_{ik} + (1 - p_d)^2 \frac{z^D f_l}{n_l^{eff}} n_{lk}\delta_{il}\\
+p_d^2 \frac{z^D f_k}{n_k^{eff}}n_{lk}\delta_{jk} +p_d(1 - p_d)\frac{z^D f_l}{n_l^{eff}}n_{lk}\delta_{jl}\\ 
+ \frac{z^N \sigma_l}{n_l}n_{lk}\delta_{il}\;.\;\;\;\;\;\;\;\;\;\;\;\;\;\;\;\;\;\;\;\;\;\;\;\;\;\;\;\;\;\;\;\;\;\;\;\;\;\;\;\;\;\;\;\;\;\;\;\;\;\;\;\;\;
\end{multline}
\begin{figure*}[t!]
	\begin{center}
	\includegraphics[width=0.88\linewidth]{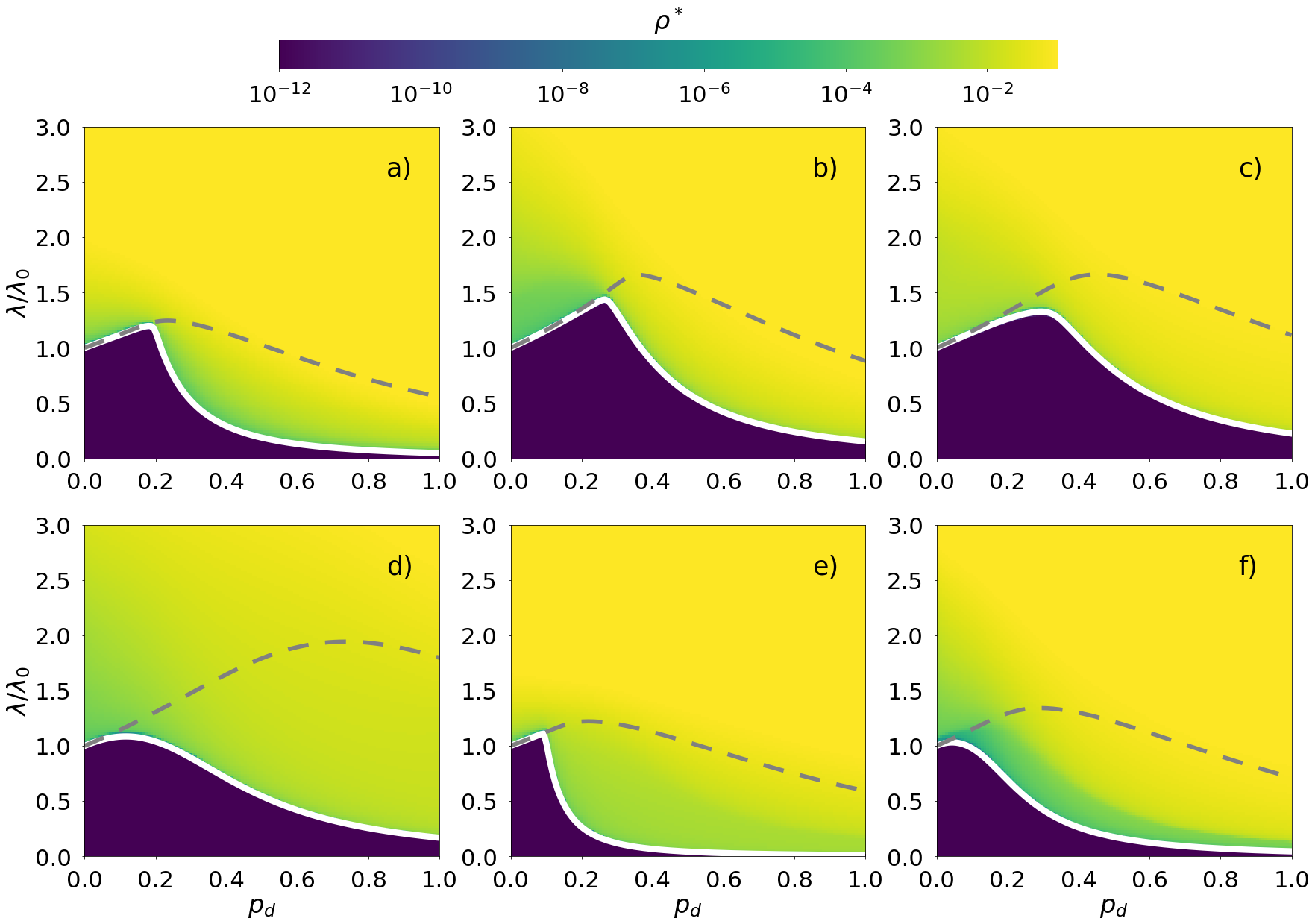}
	
%	\subfloat[]{
%		\includegraphics[height=0.26\linewidth]{"imagenes/threshold_ny"}}
%    \subfloat[]{
%		\includegraphics[height=0.26\linewidth]{"imagenes/threshold_ma"}}
%	\subfloat[]{    
%	    \includegraphics[height=0.26\linewidth]{"imagenes/threshold_austin"}}
%	\\
%	\subfloat[]{	
%		\includegraphics[height=0.26\linewidth]{"imagenes/threshold_miami"}}
%	\subfloat[]{
%		\includegraphics[height=0.26\linewidth]{"imagenes/threshold_detroit"}}
%	\subfloat[]{
%		\includegraphics[height=0.26\linewidth]{"imagenes/threshold_seattle"}}
		\end{center}
	\caption{Epidemic diagrams $\rho^*(p_d,\lambda/\lambda_0)$ for the cities of $(a)$ New York, $(b)$ Boston, $(c)$ Austin, $(d)$ Miami, $(e)$ Detroit and $(f)$ Seattle. The continuous white line shows the epidemic threshold obtained by solving Eq.~(\ref{eq:umbralDist}), whereas the dashed grey line accounts for the epidemic threshold of the indistinguishable case, {\em i.e.} obtained by solving Eq.~(\ref{eq:umbralIndist}).}
	\label{fig:heatMap}
\end{figure*}

The former matrix is the new {\em mixing matrix} for the distinguishable MIR model and, as its indistinguishable counterpart, Eq.~(\ref{mixing_indist}), captures the different ways that residents in patch $i$ traveling to $j$ mix with agents from $l$ traveling to $k$. Specifically, we can represent matrix $\bm{M}$ as follows:
\smallskip
\begin{equation} \label{eq:Matriz}
\bf{M}=
\small{
	%\frac{\mu}{\lambda} \left( \begin{matrix} \varepsilon^1_{\ 1} \\ \cdots \\ \varepsilon^1_{\ N}  \\  \_\_  \\  \cdots  \\ \_\_ \\  \varepsilon^N_{\ 1}  \\  \cdots  \\  \varepsilon^N_{\ N} \end{matrix} \right) =
	\begin{pmatrix} M^{11}_{11} & \cdots & M^{11}_{1N}| & \cdots & |M^{1N}_{11} & \cdots & M^{1N}_{1N} \\ \cdots & \cdots & \cdots & \cdots & \cdots & \cdots & \cdots \\ M^{11}_{N1} &  \cdots & M^{11}_{NN}| & \cdots & |M^{1N}_{N1} & \cdots & M^{1N}_{NN} \\ 
		\_\_\_ & \_\_\_ & \_\_\_ & \_\_\_ & \_\_\_  & \_\_\_ & \_\_\_ \\
		\cdots & \cdots & \cdots & \cdots & \cdots  & \cdots & \cdots \\
		\_\_\_ & \_\_\_ & \_\_\_ & \_\_\_  & \_\_\_ & \_\_\_ & \_\_\_ \\
		M^{N1}_{11} & \cdots & M^{N1}_{1N}| & \cdots & |M^{NN}_{11} & \cdots & M^{NN}_{1N} \\ \cdots & \cdots & \cdots & \cdots & \cdots & \cdots & \cdots \\ M^{N1}_{N1} & \cdots & M^{N1}_{NN}| & \cdots & |M^{NN}_{N1} & \cdots & M^{NN}_{NN}
	\end{pmatrix}\;.
	%\left( \begin{matrix} \varepsilon^1_{\ 1} \\ \cdots \\ \varepsilon^1_{\ N} \\ \_\_ \\  \cdots  \\ \_\_ \\  \varepsilon^N_{\ 1}  \\  \cdots  \\  \varepsilon^N_{\ N} \end{matrix} \right)
}
\end{equation}
\smallskip
Note that with the former formulation each sub-matrix $M^{il}$ of ${\bf M}$ contains the elements that relate those individuals with residence in patch $i$ with those living in $l$.\\

Taking advantage of the definition of the mixing matrix ${\bf M}$ Eq.~(\ref{eq:Linealizada}) can be written in a compact form as:
\begin{equation}
\frac{\mu}{\lambda}\vec{\epsilon}={\bf M}\vec{\epsilon}\;,
\label{eq:Linealizada2}
\end{equation}
{\em i.e} an eigenvalue problem. From all the possible solutions (eigenvectors) of Eq.~(\ref{eq:Linealizada}) we are interested in the one corresponding to the minimum value of $\lambda$ (as it defines the epidemic threshold of the metapopulation), that corresponds to the maximum eigenvalue of ${\bf M}$. Thus, the epidemic threshold reads:
\begin{equation}\label{eq:umbralDist}
\lambda_c = \frac{\mu}{\Lambda_{max}(\bf{M})}\;.
\end{equation}

\begin{figure*}[t!]
    \centering
		\label{subfig:DistvsIndist}
		\includegraphics[width=0.9\linewidth]{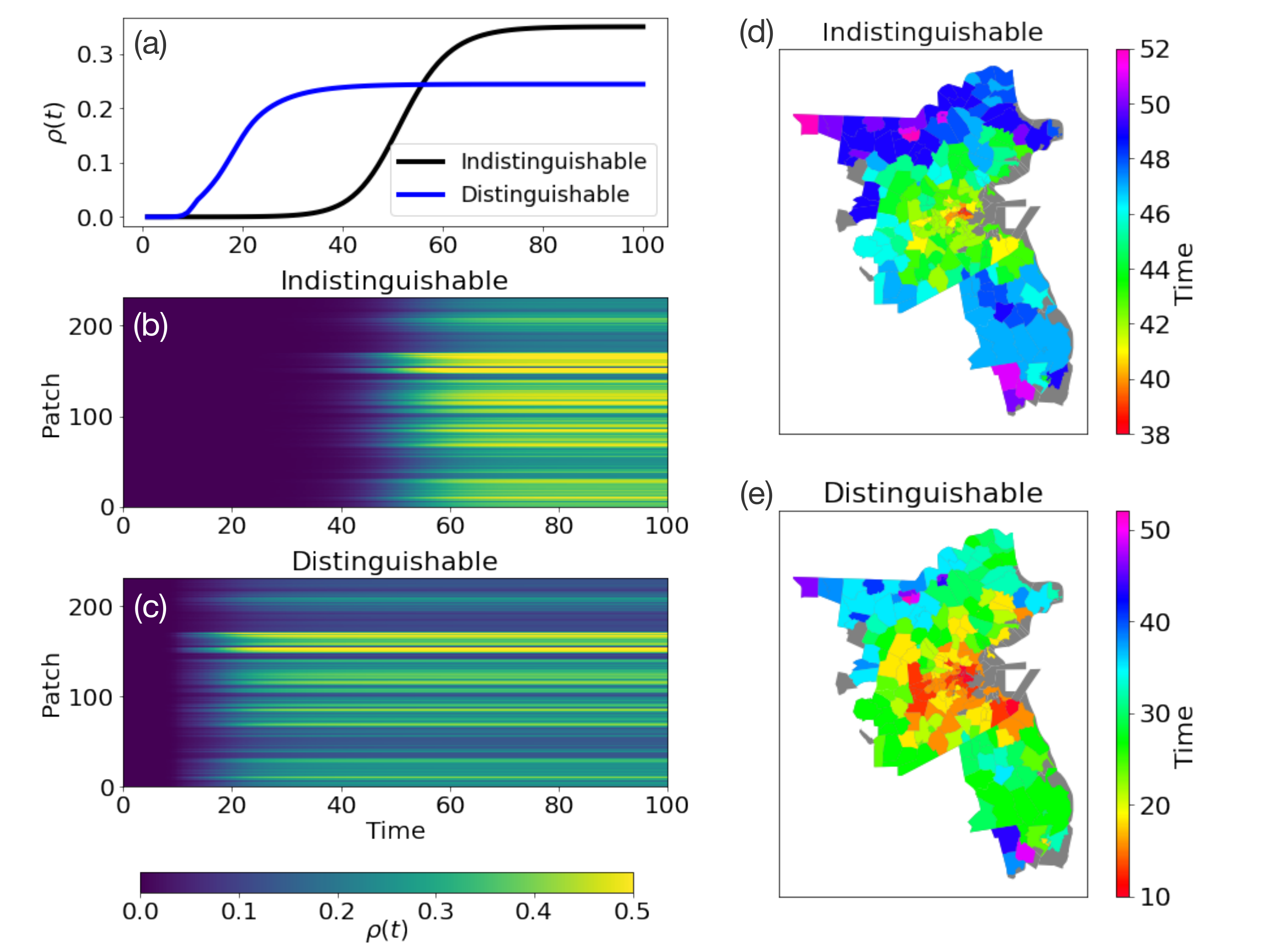}
	\caption{Spatio-temporal unfolding of an epidemic in the Boston area under the indistinguishable and distinguishable frameworks. In (a) we show the time evolution of the global prevalence $\rho^*$. Panels (b) and (c) show the time evolution of the local prevalence of each patch. Finally, panels (d) and (e) show the map of the Boston metropolitan area in which each subpopulation is colored according to the time of arrival of the first infections. This time is calculated as the time required for the instantaneous prevalence to reach $5\%$ of the local population.}
	\label{fig:4}
\end{figure*}

Finally, we verify that Eq.~(\ref{eq:umbralDist}) and the derivation of the mixing matrix (\ref{eq:mixing}) are correct by computing the epidemic diagram $\rho(p_d,\lambda)$ and comparing with the theoretical prediction for $\lambda_c(p_d)$. We have performed this analysis for $6$ US cbsa (see Table~\ref{tab:Table1} for details): New York, Boston, Austin, Miami, Detroit and Seattle. These results are shown in Fig.~\ref{fig:heatMap} where $\lambda$ has been normalized to $\lambda_0$ so that at $p_d=0$ the normalized epidemic threshold is $1$. 
In each panel we overlay to each contour plot $\rho(p_d,\lambda/\lambda_0)$ the curve $\lambda_c (p_d)/\lambda_0$ as derived by calculating the spectral radius of the mixing matrix ${\bf M}$ of each city and applying Eq.~(\ref{eq:umbralDist}). The agreement of the analytical formula is excellent and highlights the importance of considering the specific commuting patterns and the spatial distribution of the populations (the two key elements of the matrix ${\bf M}$) to assess the robustness of populations subjected to the spread of communicable diseases. In addition to the accuracy of the analytical prediction, the $6$ plots illustrate the non-monotonic trend of the epidemic threshold $\lambda_c$ as a function of mobility $p_d$ pinpointing that the epidemic detriment phenomenon remains as a generic feature for distinguishable agents as it was observed in the indistinguishable case.

\subsection{Distinguishable {\em vs.} Indistinguishable behaviors}

Although in terms of the epidemic detriment the results for indistinguishable and distinguishable agents are qualitatively similar, there are important quantitative differences between the two cases that we now analyze. First, we focus on the dependence of the epidemic threshold with mobility. In the panels of Fig.~\ref{fig:heatMap} we have included the function $\lambda_c(p_d)/\lambda_0$ (dashed grey lines) for the case of indistinguishable agents calculated as in Eq. (\ref{eq:umbralIndist}), {\em i.e} through the computation of the spectral radius of the mixing matrix for the indistinguishable case, whose elements are given by Eq~(\ref{mixing_indist}).
Let us note that the normalization factor in both curves, $\lambda_0$, remains the same for the indistinguishable and distinguishable cases since it corresponds to the epidemic threshold when no mobility is at work, $\lambda_c(p_d=0)$, which is identical for both scenarios. 

From the panels in Fig.~\ref{fig:heatMap} it becomes clear that the epidemic threshold is always smaller in the distinguishable case than in the indistinguishable one. This difference is related to the underlying flow heterogeneity of the mobility networks. In the distinguishable case, individuals who visit the main focus of infection, and thus import contagions to their residences, maintain this infectious flow between these two specific areas over time. However, in the indistinguishable case, an individual who has visited the focus may end up in a different location in the following time steps, diluting the effect of contagions between different subpopulations and thus avoiding outbreaks of secondary contagions in these subpopulations. 

These differences become more pronounced as the value of $p_d$ increases. However, when mobility is extremely low, $p_d\gtrsim 0$, the two formalisms coincide exactly and the curves $\lambda_c(p_d)/\lambda_0$ have the same slope. It is reasonable to think that, since the detriment is because the residents of the infectious focus begin to leave that node, at first order when travel is scarce the indistinguishable formalism behaves the same as the distinguishable one. In particular, for the indistinguishable case, if the interval between two consecutive trips is much longer than the duration of the infectious window of an infected individual, it will rarely visit different patches during its contagious cycle and will behave, for contagion purposes, as a distinguishable individual.

Having analyzed the differences in the epidemic threshold between the distinguishable and indistinguishable formalisms, we focus now on studying their behavior in the supercritical regime. In particular, in this regime we are interested in monitoring the transient from the initial state in which a small infectious seed is placed in a single patch to the endemic regime after the subsequent infections spread through the entire metapopulation. In Fig.~\ref{fig:4} we use as a test framework the city of Boston, and placing the same infectious seed for the two formalisms we analyze its expansion for $\lambda=2\lambda_0$ and $p_d=1$.

In Fig.~\ref{fig:4}.a we show the time evolution for the global fraction of infected individuals. It is clear that in the distinguishable case the pathogen spreads initially faster than in the indistinguishable framework. Moreover, the distinguishable prevalence reaches its stationary value when the indistinguishable prevalence is still negligible. However, once the epidemic unfolds in the indistinguishable scenario, it reaches a stationary prevalence much larger than in the distinguishable.

Apart from the different time scales involved in the transient process to the endemic equilibrium we can monitor the different spatio-temporal evolution from the initial outbreak to the stationary regime. The two epidemic trajectories are compared in Figs.~\ref{fig:4}.b-c. From these two plots it is clear that the epidemic unfolding under the two frameworks follows different paths although, given the different times scales involved, it is difficult to pinpoint the localization of these differences. To shed more light on the different spatio-temporal patterns of each propagation process, in Figs.~\ref{fig:4}.d-e we have drawn the map of the Boston area by assigning each patch a time value. This value accounts for the exact time at which the infected fraction of the patch reaches $5\%$ of its population. It is clear from the two colored maps that the non-distinguishable case follows a more explosive behavior, so that once the contagions spread out of the central Boston area, where the infectious seeds are located, the disease reaches the entire metropolitan population in a few time-steps. In contrast, for the distinguishable frame the process is highly sequential and takes more than $40$ time steps to reach the entire population following, moreover, a radial pattern from the center to the periphery.

\section{Conclusions}
\label{sec:conc}

In this paper we have developed a Markovian framework based on the MIR model to analyze the spread of communicable diseases in networked metapopulations. Unlike its original formulation in which agents are labeled according to their residence, the new framework incorporates the distinguishable nature of agents according to both their residence and recurrent destinations. This new framework incorporates a further partition of subpopulations that can be easily gathered from commuting data and is certainly valuable for studying the spread of diseases in metropolitan areas since recurrent paths apply to most of the mobility flows. 

After validating the Markovian equations by comparing with results obtained with mechanistic agent-based simulations, we have derived a new mixing matrix that captures the basic interaction mechanisms between the subpopulations of agents at different patches. By calculating the spectral radius of this mixing matrix we can estimate the epidemic threshold of the metapopulation, allowing us to estimate its robustness to the spread of pathogens. 

Finally, we have compared the results obtained considering indistinguishable and distinguishable agents, taking advantage of the fact that, in the limit of null mobility, both frames are identical. Firstly, we have shown that indistinguishability overestimates the value of the epidemic threshold, demonstrating how distinguishability does not allow contagions produced in areas of high prevalence to be diluted between different patches, making it easier for them to cause secondary infections. This overestimation of the epidemic threshold, however, does not eliminate the phenomenon of the epidemic detriment that is still observed in the case of distinguishable agents. Likewise, we have observed that the spatio-temporal diffusion for the distinguishable case occurs more progressively than in the indistinguishable case, spreading spatially much more explosively.

Our manuscript fuels the discussion on the relevance of the nature of the mobility schemes introduced in the theoretical framework to provide a fair assessment of the evolution of diseases. In random-walker dynamics, Castioni et al.~\cite{castioni2021critical} demonstrate that the outcome of control policies shaping mobility is strongly shaped by the ratio between the time scales involved in both movements and contagions. This control parameter is also crucial when one accounts for recurrent mobility patterns and its variation changes the critical properties of the metapopulation, leading to a vanishing of the epidemic detriment here reported in some scenarios~\cite{JSTAT}. On more general grounds, different theoretical works have shown that the information loss when translating higher-order flows in origin-destination matrices~\cite{matamalas2016assessing}, the introduction of biases in the collection of mobility data~\cite{tizzoni2014use,schlosser2021biases} or the misuse of raw mobility data as OD matrices~\cite{gomez2019impact} leads to substantial differences in the evolution of epidemics.

Despite its simplicity, the new distinguishable framework of the MIR model opens the door to the implementation of more accurate descriptions of real urban environments, a context where recurrent mobility flows predominate and  precise identification of possible contagion pathways is much needed. Nevertheless, it is worth stressing that the model here proposed does not capture entirely the weekly mobility rhythms of the population, as assuming a fixed destination over the weekends misrepresents the heterogeneous and variable nature of our usual mobility patterns in leisure time~\cite{alessandretti2018evidence,schlapfer2021universal}. The formulation of a framework taking into account the time-varying nature of our mobility patterns remains as future work. In epidemiological terms, we have focused here on a simplified version in which an epidemic SIS model is at work, but this formalism can be generalized to any other compartmental dynamics along with other refinements such as considering the age partitioning of the population or the inclusion of more complex interactions and mobility patterns.

\section*{Data Availability Statement}
The data that support the findings of this study are available from the corresponding author upon reasonable request.

\section*{ACKNOWLEDGEMENTS \label{sec:acknowledgements}}
P.V., D.S.P. and J.G.G. acknowledge financial support from the Departamento de Industria e Innovación del Gobierno de Aragón y Fondo Social Europeo (FENOL group E-19), from grant PID2020-113582GB-I00 funded by MCIN/AEI/10.13039/501100011033,  and from Fundación Ibercaja and Universidad de Zaragoza (grant 224220). A.A. acknowledges financial support from Spanish MINECO (Grant No.\ PGC2018-094754-B-C21), Generalitat de Catalunya (Grant No.\ 2017SGR-896 and 2020PANDE00098), Universitat Rovira i Virgili (Grant No.\ 2019PFR-URV-B2-41), Generalitat de Catalunya (PDAD14/20/00001), ICREA Academia, and the James S.\ McDonnell Foundation (Grant No.\ 220020325).

%\section*{REFERENCES}
%\nocite{*}
\bibliography{References}% Produces the bibliography via BibTeX.
% \begin{thebibliography}{10}

\end{document}